\documentclass[a4paper]{ifacconf}

\usepackage{graphicx,amsmath,url}      % include this line if your document contains figures
\usepackage[round]{natbib}             % required for bibliography
\usepackage{caption}
\ifx\portugues\undefined
\def\portugues{0}
\fi

\if\portugues0
   \usepackage[english]{babel}
  \else
   \usepackage[spanish,brazil,english]{babel}
\fi

\usepackage[T1]{fontenc}

\usepackage[utf8]{inputenc}

\usepackage{ae}

\if\portugues1
 { }
 { }
 { }
 
\fi

\begin{document}

\if\portugues1

% =====================================================================
% =====================================================================
% USE THIS PART IF THE TEXT IS IN PORTUGUES OR SPANISH
% =====================================================================
% If the manuscript is in Spanish, please change the texts adequately.
% =====================================================================
% 
\selectlanguage{brazil}
	
\begin{frontmatter}

\title{Sistema Embarcado para Registro e Controle de Higienização de Mãos em Ambientes de Saúde
%\thanksref{footnoteinfo}
} 

%\thanks[footnoteinfo]{Reconhecimento do suporte financeiro deve vir nesta nota de rodapé.}

\author[First]{Rafael Z. Castro}
\author[First]{Alexandre S. Roque}

\address[First]{Curso de Engenharia Elétrica, Universidade Regional Integrada do Alto Uruguai e das Missões, Campus Santo Ângelo, RS, (e-mail: rafaelzcastro@aluno.santoangelo.uri.br, roque@santoangelo.uri.br).}
%\address[Third]{Electrical Engineering Department, 
%   Seoul National University, Seoul, Korea, (e-mail: author3@snu.ac.kr)}

\selectlanguage{english}
\renewcommand{\abstractname}{{\bf Abstract:~}}
\begin{abstract}                % Abstract of not more than 250 words.
Hand hygiene (HH) control is crucial to prevent cross-contamination and healthcare-associated infections (HAI), according to Brazilian regulatory standards and WHO guidelines. The present study addresses the lack of automation technologies for HH, aiming to record, measure, and provide data for internal audits in hospitals. This article introduces an embedded system for HH control and recording, comprising low-cost hardware architecture with IoT connectivity and online monitoring. Results with practical evaluation in a real hospital setting for 3 hours demonstrated the system's effectiveness in recording HH indices.

\vskip 1mm% não altere esse espaçamento
\selectlanguage{brazil}
{\noindent \bf Resumo}:  O controle da higienização de mãos (HM) é essencial para evitar contaminações cruzadas e as infecções relacionadas a assistência à saúde (IRAS), de acordo com Normas da ANVISA e da OMS. A motivação deste estudo emerge da carência de tecnologias de automação deste processo de HM, permitindo registrar, mensurar, e disponibilizar esses dados para auditoria interna de ambientes hospitalares. Assim, o presente artigo apresenta um sistema embarcado para controle e registro dos indices de HM, composto por uma arquitetura de hardware de baixo custo com conectividade IoT, e um ambiente online de monitoramento. Resultados práticos em cenário de teste real durante 3 horas em ambiente hospitalar, mostraram a efetividade do sistema ao registrar os índices de higienização.   
\end{abstract}

\selectlanguage{english}


\selectlanguage{brazil}

\end{frontmatter}
\else
% ===============================================================
% ===============================================================
% USE THIS PART IF THE TEXT IS IN ENGLISH
% ===============================================================
% ===============================================================
% 

\begin{frontmatter}

\title{Embedded System for Recording and Controlling Hand Hygiene in Healthcare Environments\thanksref{footnoteinfo}} 

\thanks[footnoteinfo]{Thanks to Hospital Unimed from Santo Ângelo city, RS, Brazil, for providing the testing environment, providing \textit{feedback} for future improvements to the prototype.}

\author[First]{Rafael Z. Castro}
\author[First]{Alexandre S. Roque}

\address[First]{Electrical Engineering Department, Regional Integrated University of High Uruguai and Missions, Santo Angelo Campus, RS State, (e-mail: rafaelzcastro@aluno.santoangelo.uri.br, roque@santoangelo.uri.br).}
   
\renewcommand{\abstractname}{{\bf Abstract:~}}   
   
\begin{abstract}  % Abstract of not more than 250 words.
Nowadays, more effective control of hand hygiene (HH) by healthcare teams has become essential. HH control is crucial to prevent cross-contamination and healthcare-associated infections (HAI), according to Brazilian regulatory standards and WHO guidelines. The lack of widespread technology to measure acceptable hygiene rates within hospital environments leads to the practice of a manual sample audit reading, requiring more time for decision-making. Thus, the present study addresses the lack of automation technologies for HH, aiming to record, measure, and provide data for internal audits in hospitals. This article introduces an embedded system for HH control and recording, comprising low-cost hardware architecture with IoT connectivity and online monitoring. Results with practical evaluation in a real hospital setting for 3 hours demonstrated the system's effectiveness in recording HH indices.
\end{abstract}


\end{frontmatter}
\fi

%===============================================================================
%===============================================================================
%===============================================================================

\section{Introduction}

The COVID-19 pandemic has highlighted the importance of hand hygiene (HH) as a crucial measure in preventing the spread of infections. Healthcare professionals must perform HH diligently to prevent the transmission of pathogens, bacteria, and fungi. However, the routine of these environments makes it difficult to effectively monitor HH, which is generally carried out through periodic audits conducted by the Hospital Infection Control Service (HICS). Currently, these audit processes are guided by health surveillance sectors, but sample data collection can lead to distortions in numbers. 

The HICS conducts audits and analyzes data to obtain an HH effectiveness rate, classified by area and type of health professional. When this rate is lower than the desired indexes, the team promotes awareness and inspection actions. Therefore, the proposal for an automated system that facilitates the measurement of the hand hygiene rate is relevant, as there is a lack of technological development in an embedded system to measure and raise awareness among healthcare teams about hand hygiene \citep{de2020importancia}.

A relevant aspect in the current technique for measuring hand hygiene rates is the \textit{Hawthorne} effect, which consists of the posture of health professionals, who behave differently when they know that have been observed. This posture can mask the data, bringing a false reality of correctness in the data \citep{purssell2020hawthorne}.

According to the latest WHO assessment of the National Program for the Prevention and Control of healthcare-associated infections – PNPCIRAS, item 6.2.1, monitoring and feedback on hand hygiene compliance is identified as a key national indicator, citing the lack of feedback, i.e., considering no proven effectiveness \citep{OMS2022}.

In hospital environments, RFID and IoT technologies have demonstrated efficiency in reducing the impacts caused by HAI. Whether strengthening contagion prevention measures or improving methods of care for infected people, these technologies prove to be effective \citep{costa2022aplicaccao} \citep{subrahmannian2022chipless}.

Thus, this work presents the development of an embedded system prototype to control hand hygiene automatic devices and measure their effectiveness rate. The application proposal brings several benefits, creating standardization in obtaining data between the different areas of the Health Care Institution (HCI). The embedded system is a starting point for application development and improved sensing in this type of application. Thus, automation and engineering technologies emerges as a promising solution for addressing the problem of HH monitoring.

This article is structured as follows: Section 2 presents a brief theoretical basis with related work and a description of the problem; Section 3 describes the methodological procedures; Section 4 specifies the development and architecture of the embedded system prototype; Section 5 discusses validation scenarios, tests, and results; Finally, Section 6 presents the conclusions and future perspectives of the study.

\section{Related Work and Problem Statement}

\subsection{Related legislation}

In Brazil, ANVISA established RDC nº 42/2010, which provides guidelines for hand hygiene in health services. ANVISA's RDC No. 50/2002 includes mandatory washbasins \citep{BRASIL2002} \citep{BRASIL2010}. According to WHO guidelines, the total duration of the hand hygiene procedure should be 40 to 60 seconds \citep{santos2021identification}.

\subsection{Related Work}
Based on the current legislation mentioned, it is possible to use technologies that aim to automate the hand hygiene process, enabling constant monitoring and assisting health units. Thus, some related works are presented.

According to \cite{queiroz2021aplicaccao}, the use of embedded systems with RFID and/or IoT has proven to be an efficient strategy in hospital environments, as it allows the tracking and monitoring of various aspects related to infection control, resulting in minimized impacts.
In \cite{li2018wristwash} the project \textit{WristWash} consists of monitoring the HH technique, using a bracelet with an IoT device, which contains accelerometers and identifies the different moments within the hand hygiene technique, comparing with patterns the result of the accelerometer coordinates.

In the \cite{cherin2018hygiene} work, hand hygiene techniques are monitored with the support of Artificial Intelligence algorithms, specifically with machine learning, where a camera recognizes the different moments of hand hygiene. Another approach is presented in \cite{bal2017system} with cloud-connected stations that can cooperate to detect hand hygiene events in real-time, but reports are not online and do not consider different HH opportunities.

In \cite{fagert2022clean} \textit{CleanVibes} is proposed, which addresses a technique for monitoring hand friction during hand hygiene, which, through the application of the Fast Fourier Transform, obtains patterns, which identify the different moments within the hygiene technique.
Previous approaches are focused on monitoring handwashing action. On the other hand, the \cite{wu2020autonomous} proposal focuses on autonomously registering the healthcare professional, using \textit{Bluetooth Low energy} (BLE) and IoT, with each professional having a wearable device with their functional ID, to register when it is close to the washing dispenser. However, the work requires checking other people entering the beds, in addition to using more complex hardware.

The need for an embedded system to measure the rate of hand hygiene, with runtime data, contributes to infection prevention. There is a lack of devices for this purpose, due to the emerging demand for hospital infection control services. In hospital environments, embedded systems have demonstrated efficiency in different automation processes, adding value and strengthening contagion prevention measures or improving methods of care for infected people \citep{krishnamoorthy2023role}.  The Table \ref{tb:RWSummany} summarizes the related work considering the current research goal.

\begin{table}[ht]
\begin{center}
\caption{Related Work summary.}\label{tb:RWSummany}
\begin{tabular}
{p{1.8cm}p{3.1cm}p{2.7cm}}
Ref: & Main approach & Gap \\\hline
\cite{queiroz2021aplicaccao} & IoT and RFID in hospitals allowing the infection monitoring & Not consider HH \\
\cite{li2018wristwash} & HH monitoring with a bracelet IoT & Not consider opportunity control \\
\cite{cherin2018hygiene} & HH monitoring with AI and cameras & Not consider opportunity control\\
\cite{bal2017system} & Cloud-connected hand hygiene stations & Not consider opportunity control\\
\cite{fagert2022clean} & Hand friction monitoring to obtain patterns in HH & Online and opportunity control \\
\cite{wu2020autonomous} & BLE and IoT wereable device on professionals & Complex and not consider opportunity control \\
\cite{krishnamoorthy2023role} & Overview about IoT and Hospital 4.0 & Highlight different technologies but lack HH opportunities.\\ \hline
\end{tabular}
\end{center}
\end{table}

The works presented have their specificities, related to the application environment and sensory technologies focused on verifying the hygiene procedure. The problem addressed in this work is related to the lack of onboard devices to monitor ``hand hygiene opportunities'' when entering and leaving hospital beds, especially in Intensive Therapy Units (ITUs). The present study differs mainly by controlling the opportunity when the employee or visitor arrives at the location, with an educational illustration of the cleaning steps, a historical record of process data, and connectivity with an IoT cloud environment, in addition to a control algorithm embedded in a device, portable, which can be attached to ITU washbasins and regular beds. 

\section{Methodological procedures}

\begin{figure}[!ht]
\begin{center}
\includegraphics[width=0.8\linewidth]{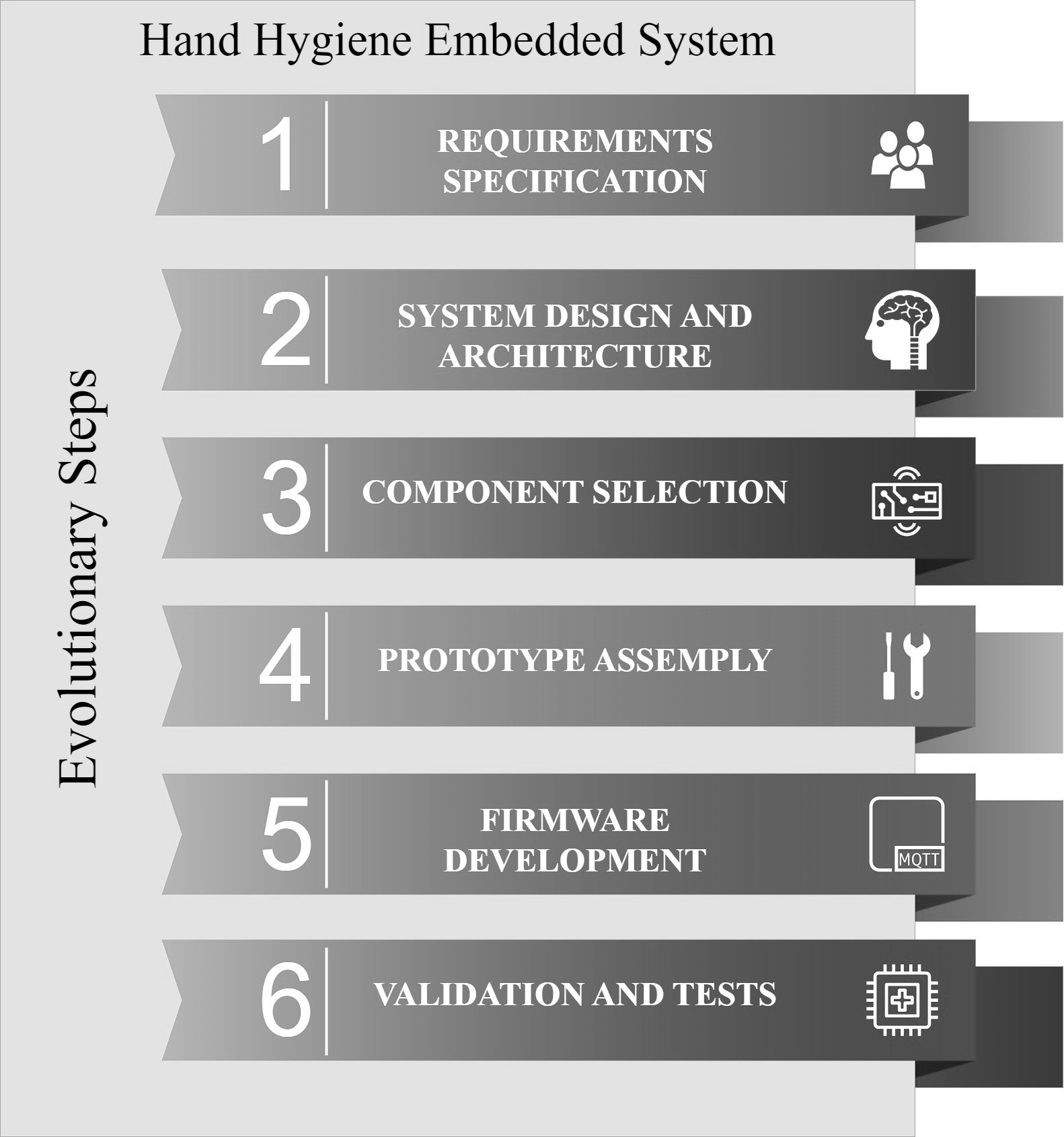}    % The printed column width is 8.4 cm.
\caption{Steps in the Methodological procedures.} 
\label{fig:procmetodologicos}
\end{center}
\end{figure}

The research and experimentation process adopted in the present study follows the premises addressed in \cite{azevedo2020metodologia} and \cite{martelli2020analise}. Figure \ref{fig:procmetodologicos} illustrates the steps adopted in the research, characterizing an evolutionary prototyping method.

Regarding the type of research, the work is classified as applied research, due to the practical emphasis on solving a specific problem and developing an embedded system that automates the process of monitoring and controlling the rate of hand hygiene in a hospital environment. In terms of nature, this study is characterized by being qualitative research due to the specific proposal for a solution to the problem in question. As for its purposes, it is characterized by being descriptive and explanatory, through the possibility of automating a process that is currently carried out by verification and sample audits, where the current result is inaccurate due to the \textit{Hawthorne} effect. As for the means, it is characterized by being a bibliographical, experimental, and laboratory research, with the development of a prototype, to be a starting point for embedded systems for HH in health institutions.

\section{Embedded System for Hand Hygiene}

For the embedded system development, a survey of needs and requirements was carried out in conjunction with HICS (Hospital Infection Control Service). The need to include guidance on the necessary steps for HH was observed, in a clear and easy-to-understand format for the user. 

\subsection{Requirements Specification}
Therefore, the main requirements specified were: Avoid physical contact to activate water; Request hand hygiene when accessing the bed; Request hygiene after accessing the bed; Prototype developed with smooth surfaces to prevent bacterial spread; Conduct formal hand hygiene guidance in conformance with the WHO and HCI recommendations where equipment is tested; Calculate the hand hygiene rate during operation time; Provide the opportunity to identify the professional who carried out the hand hygiene.

\subsection{Embedded System Architecture and Components}
\begin{figure}[!ht]
\begin{center}
\includegraphics[width=0.7\linewidth]{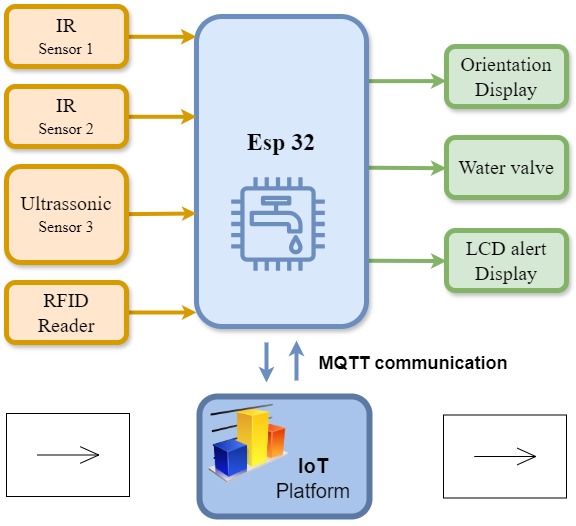}    % The printed column width is 8.4 cm.
\caption{Proposed Architecture Implemented.} 
\label{fig:diagramaARQ}
\end{center}
\end{figure}

Based on the specified requirements, an interconnection architecture for the electronic and software components that integrate the embedded system was structured. Additionally, the system communicates with an IoT platform that was chosen for the project. In an illustrative way, the block diagram in Figure \ref{fig:diagramaARQ} illustrates how the components interact with the microcontroller and the IoT platform for data visualization, the \textit{Losant} (www.losant.com). 

The embedded system has four general elements: those responsible for reporting input data (sensors), the output devices (three elements composed of an LCD display, an orientation display and a water actuation valve), the processing unit (ESP32) and the IoT monitoring platform.

Composing the architecture and connection needs, the components were integrated with the ESP32 microcontroller, which is widely used in IoT projects. The inputs are composed of sensors 1 and 2, which are infrared sensors model GP2Y0A21YK0F from Sharp (80cm range), used to detect the entry and exit of a person in the hospital bed, characterizing HH opportunities. The ultrasonic sensor model HC-SR04 is applied in front of the tap, to detect presence in front of it. The 13.56 MHz RFID reader model MFRC522 is used for optional identification of the person/professional at the entrance to the bed. The outputs are presented in two ways, with the orientation display (an innovative differentiator of the present work), a 16x2 LCD display for information on the cleaning steps, and the tap control actuator, a 12V solenoid valve for 1/2 x 1/2.

\begin{figure}[!ht]
\begin{center}
\includegraphics[width=1.0\linewidth]{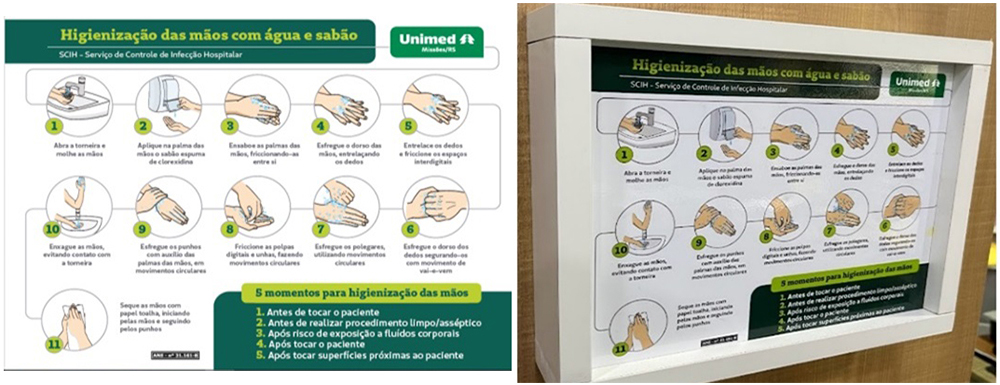}    % The printed column width is 8.4 cm.
\caption{Orientation Display of the HH Process.} 
\label{fig:displayO}
\end{center}
\end{figure}

One of the differences of this work was the use of a display guiding the 11 hygiene steps (according to WHO and HCI). These steps are illustrated in Figure \ref{fig:displayO} and below the designed and assembled display, containing rear LEDs backlighting, in a timed manner, each step whenever a person starts the HH process in front of the tap.

After this step, to accommodate the electronic components, a cabinet was designed for the physical installation of the device in a healthcare environment. For the cabinet construction, acrylic was chosen, which is a smooth and easy-to-clean product, as verified in the requirements specification. Thus, a \textit{Case-Box} was made where the view of the display is at an ergonomic angle to the user's view, thus housing the circuit board, LCD display, ultrasonic sensor, and RFID reader module. Both ``Case-Box'' and ``Orientation Display'' components were installed in a piece of furniture designed for this application. A basin, tap, and fixture were dedicated to affixing the display, thus allowing the controller and other electronics to be placed in a single piece of furniture.

\subsection{Prototype Assembly}
\begin{figure}[!ht]
\begin{center}
\includegraphics[width=1.0\linewidth]{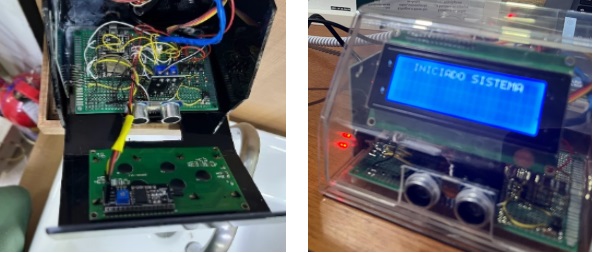}    % The printed column width is 8.4 cm.
\caption{Circuit assembly and Case-Box installation.} 
\label{fig:montagemProt1}
\end{center}
\end{figure}

Assembling the prototype, calibration, and measurement procedures for the sensors were carried out in the laboratory, certifying the minimum distance of one meter for the infrared sensors, ensuring detection at the appropriate distances, as well as verifying the correct reading of the RFID cards in the designed Case-Box. The electronic circuit for interconnecting the embedded system components, represented in Figure \ref{fig:diagramaARQ}, was mounted on a perforated plate, improving signal conduction. Such assembly and installation are shown in Figure \ref{fig:montagemProt1}.

\begin{figure}[!ht]
\begin{center}
\includegraphics[width=0.8\linewidth]{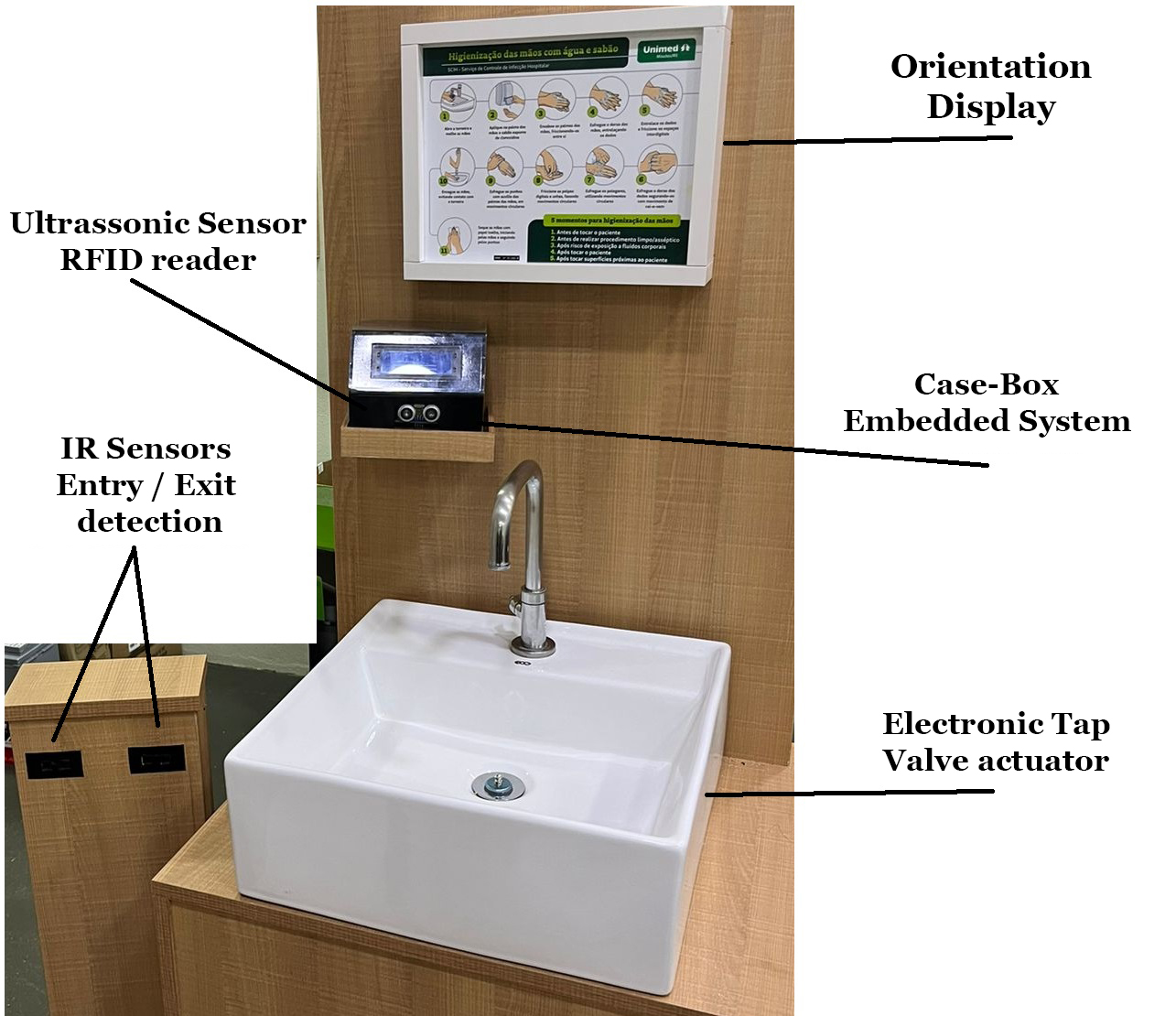}    % The printed column width is 8.4 cm.
\caption{Final prototype assembly.} 
\label{fig:montagemProt2}
\end{center}
\end{figure}

Following the methodological procedures, the components were then fixed and arranged in the Case-Box and on the designed furniture, finishing with a black adhesive coating for finishing, as shown in Figure \ref{fig:montagemProt2}.

With the prototype assembled, functional tests, RFID communication, and verification of minimum distances to detect people entering and leaving began, with the furniture with the washbasin positioned at the bed entrance. Such functional tests are fundamental for parameterizing the algorithm, which is presented in the next section.

\subsection{Firmware Development}

Research was conducted analyzing similar features and products for the purpose of this work, as presented in the research by \cite{wu2020autonomous}. Thus, to perform control in conformance with the requirements raised, the control algorithm was embedded in the ESP32 platform, with WiFi communication capabilities (sending data to the IoT platform using the MQTT protocol), I2C for the display, and SPI for the RFID reader. The control technique adopted is characterized as a closed loop approach, where the algorithm is fed back by a constant monitoring function of the sensors, providing feedback for starting, stopping, and processing the HH registration, in addition to the user identification.

Monitoring of the 11 HH stages only occurs after re-reading the distance sensor to validate that the user is in front of the embedded system (in front of the tap), if it is not within the preliminarily configured distance, it aborts and returns to the beginning not validating the HH. In other words, to validate an HH, the user needs to remain in front of the tap throughout the timed process of the 11 HH stages, which is visible in an educational way on the orientation display.

The control algorithm is divided into 3 general steps: in step 1, all programming libraries are imported and configured, which simplifies the development; In step 2, all control variables are specified, with emphasis on the number of ``hospital bed accesses'', number of ``bed exits'', ``number of occupants'', ``number of HH opportunities'', ``HH rate'' and ``number of HH''; Step 3 specifies the main control modules, including Setup (general I/O configuration), Loop (general algorithm operation function), Enter (identification of people entering the bed), Exit (identification of people leaving of the hospital bed), ConnectAutomatedTap (connection via MQTT of the embedded system with the Losant IoT platform), MonitorInputOutput (constant monitoring of sensors), Distance (measurement of the distance between user and the embedded system), PublishData (sending data to the Losant IoT platform), Procedure11 (verification and control of the 11 steps of the guidance display), and Abort (cancels ongoing operation when the user interrupts the process in the middle of HH).

\begin{figure*}
\begin{center}
\includegraphics[width=0.9\linewidth]{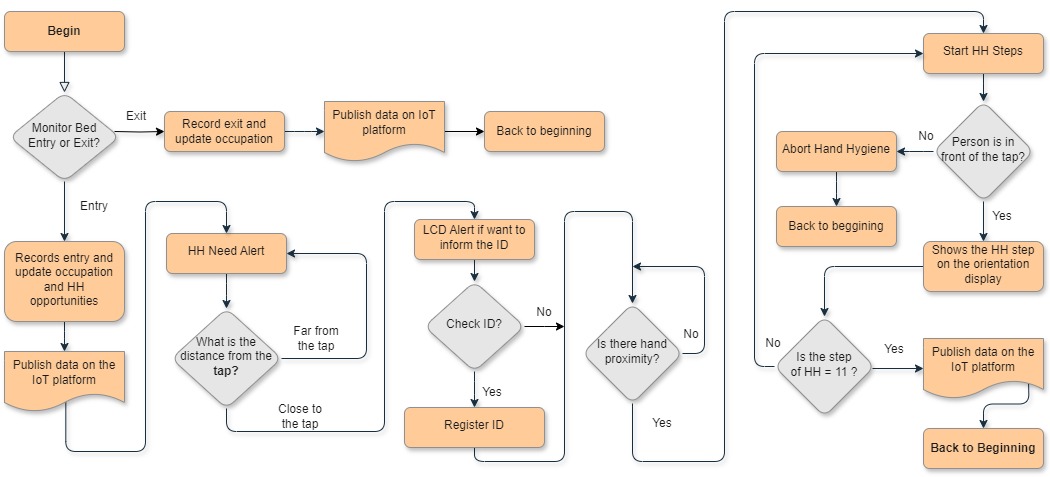}    % The printed column width is 8.4 cm.
\caption{General flowchart of the control algorithm.} 
\label{fig:diagramaAlgoritmo}
\end{center}
\end{figure*}

All these functions define a modular programming model that facilitates the maintenance and identification of all procedures coded in the embedded system control. The complete source code is available in full on GitHub via the link https://github.com/zavalik1986/Torneiraiot/. Figure \ref{fig:diagramaAlgoritmo} presents the logical sequence of execution and decision making of the control algorithm.

A relevant challenge planned was access and exit control, as each person can have a different speed when walking, between 1m/s and 3m/s, on average. Based on the flowchart presented in Figure \ref{fig:diagramaAlgoritmo}, the algorithm monitors people's access and, when it occurs, adds two HH opportunities, simultaneously recalculating the environment occupancy (in the hospital bed). If there is occupancy, the system waits for the tap to approach and displays the need for hand hygiene on the display, starting the 11-step process. The user is offered the identification option, carried out by placing the badge close to the RFID sensor on the left side of the display. This option is not mandatory, but, when used, it captures the hexadecimal code of the TAG at 13.56 MHz. Upon completing the 11 steps, the system recalculates the hand hygiene rate and publishes the data on the Losant IoT platform.

\section{Hospital Validation and Results}

During the test in a hospital environment, the embedded system was installed for a period of 3 hours, where it was possible to observe in practice the effectiveness of the system application. The test was performed in a real environment, at the entrance to a hospital bed, where the institution's management team validated the system. Practical testing procedures in the hospital environment were adopted, where real-time hand hygiene opportunities and rates were effectively recorded. 

Table \ref{tb:testTab} presents some data collected on a sample basis (48 min), to exemplify the effectiveness of the prototype in its practical tests.

\begin{table}[hb]
\begin{center}
\caption{Data sample collected in the tests.}\label{tb:testTab}
\begin{tabular}{ccccccc}
Timestamp & TX Hyg & NO & NS & NAc & NE & NOc \\\hline
%21/11/2023 09:32:00 & 50,00\% & 2 & 1 & 1 & 0 & 1 \\
21/11/2023 09:44:40 & 100,00\% & 2 & 2 & 1 & 0 & 1 \\
21/11/2023 09:57:25 & 75,00\% & 4 & 3 & 2 & 1 & 1 \\
21/11/2023 09:59:38 & 33,33\% & 12 & 4 & 6 & 4 & 2 \\
21/11/2023 10:28:59 & 31,25\% & 16 & 5 & 8 & 5 & 3 \\
21/11/2023 10:30:56 & 37,50\% & 16 & 6 & 8 & 5 & 3 \\
21/11/2023 10:32:43 & 38,89\% & 18 & 7 & 9 & 5 & 4 \\ \hline
%21/11/2023 11:01:25 & 25,00\% & 4 & 1 & 2 & 1 & 1 \\
%21/11/2023 11:08:16 & 16,67\% & 12 & 2 & 6 & 4 & 2 \\ \hline
\end{tabular}
\end{center}
\end{table}

Where TX Hig: Hand hygiene rate, NO: Number of opportunities, NS: Number of sanitizations, NAc: Number of accesses, NE: Number of exits, NOc: Number of occupations.

\begin{figure}[!ht]
\begin{center}
\includegraphics[width=0.5\linewidth]{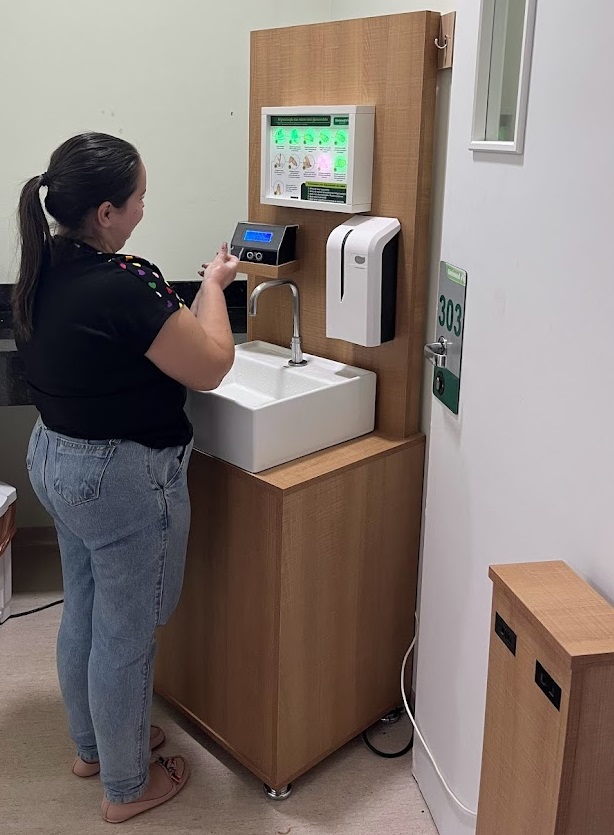}    % The printed column width is 8.4 cm.
\caption{Test Recording at the hospital bed entrance.} 
\label{fig:validacao1}
\end{center}
\end{figure}

The data in Table \ref{tb:testTab} was recorded during initial tests and collected directly from the data output on the Losant IoT platform. The System presented some instability problems due to a drop in internet connection, but it continued to function normally after the connection was restored. The figure \ref{fig:validacao1} illustrates one of the testing moments with Hospital employees.

\begin{figure}[!ht]
\begin{center}
\includegraphics[width=1.0\linewidth]{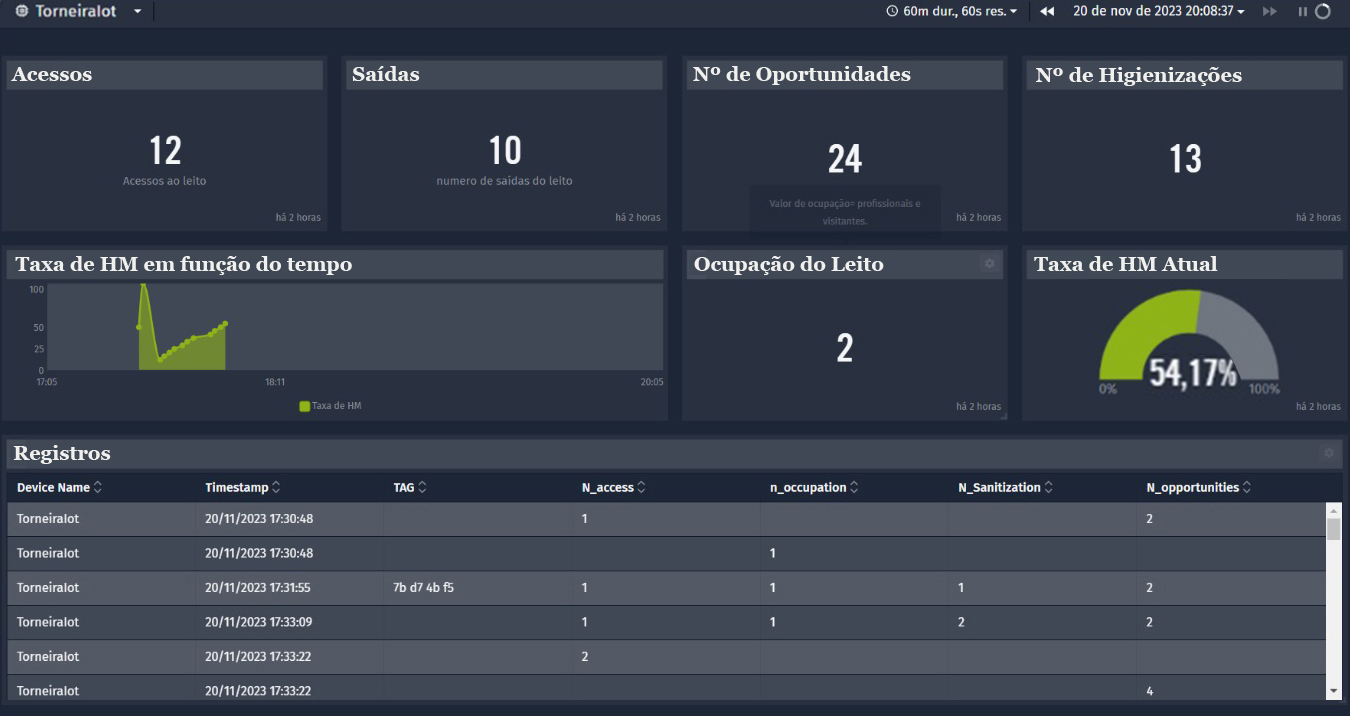}    
\caption{Hand Hygiene monitoring system screen.} 
\label{fig:dashboardVisual}
\end{center}
\end{figure}

After validation and testing in a hospital environment, the embedded system proved to be an effective tool for monitoring hand hygiene. During the bed entry and exit steps, the infrared sensors proved to be efficient, however, the people flow must be controlled. All recorded data was structed in ``data points'' and sent to the IoT Losant platform, with a specific \textit{dashboard} for better visualization. Figure \ref{fig:dashboardVisual} shows the hand hygiene monitoring system screen configured and in operation. The screen highlights the number of hospital bed accesses (12), exits (10), the number of opportunities (24), the number of HH (13), and bed occupancy (2), in addition to 3 status graphs, with emphasis on the percentage of HH at 54.17\%.

%\begin{figure*}
%\end{figure*}

This \textit{dashboard} contains eight blocks with data on the HH rate status over time. The first block shows the ``number of accesses'' variable, which means, it shows all detected environment accesses. The second block demonstrates the variable ``number of exits'' from the environment. The third block demonstrates the ``HH opportunities'' generated by the embedded system. The fourth block shows the ``number of complete sanitizations'' carried out. The fifth block graphs the ``HH rate over time'', which can be adjusted by the user. Thus, it is possible to monitor moments of lower adherence to HH practice. The sixth block shows the ``number of occupants'' in the room up to that moment. The seventh block graphically demonstrates the ``current HH rate''. The eighth block is the lower table that contains the records depending on time, being able to observe, in addition to the previous data, the identification of the professional through the hexadecimal RFID code.

Finally, an important aspect is maintaining the internet connection, which must remain stable. In the event of power outages, there was a need to keep data in internal memory, but in the future recording on an SD card could be implemented. Data is sent to the Losant platform after each HH process is completed. By monitoring proper compliance with hand hygiene practices through RFID badges, it is possible to increase healthcare professionals' adherence to these essential guidelines.

\section{Conclusions}

The development of this embedded system contributed to the improvement in the process of controlling HH rates in the hospital environment, being a more precise mechanism, avoiding the \textit{Hawthorne} effect, with the data stratification in any period and in operation time, where currently, this task is conducted through sample observation. Another fundamental contribution was the connection of the system with the IoT Losant platform, adding data visualization benefits for HCI managers, who can monitor the evolution of hand hygiene rate indicators. The scalability capacity is another important contribution, as the embedded system can be applied in different areas of the HCI, with more devices, further improving the control of HH rates, including Intensive Care Units and Surgical Centers, in addition to other critical areas. 

During the validation period, the institution's employees were subjected to an evaluation questionnaire, where \textit{feedback} was recorded that the procedures met the standards, with correct operation and information recorded correctly during the HH process. Automation systems in healthcare environments, with connectivity, drive the connection of what we call Industry 4.0 to the inclusion of the Hospital 4.0 concept, as healthcare establishments still have many opportunities for supporting technological development, as it is known that Robotics and Automation are already included in various equipment, including current robotic surgeries.

Future work considers the insertion of machine learning methods, to guide HH steps, with supervision of hand movements, supported by a camera added to the embedded system; IoT platform customization or one developed specifically for this project; and insertion of records on memory cards for periods of connection instability.

%\section*{Agradecimentos}
%Agradecimento ao Hospital Unimed de Santo Ângelo, RS, por disponibilizar o ambiente de testes, proporcionando \textit{feedback} para melhorias futuras no protótipo. 

\bibliography{ifacconf}             % bib file to produce the bibliography
                                                     % with bibtex (preferred)
                                                   
%\begin{thebibliography}{xx}  % you can also add the bibliography by hand

%\bibitem[Able(1956)]{Abl:56}
%B.C. Able.
%\newblock Nucleic acid content of microscope.
%\newblock \emph{Nature}, 135:\penalty0 7--9, 1956.

%\bibitem[Able et~al.(1954)Able, Tagg, and Rush]{AbTaRu:54}
%B.C. Able, R.A. Tagg, and M.~Rush.
%\newblock Enzyme-catalyzed cellular transanimations.
%\newblock In A.F. Round, editor, \emph{Advances in Enzymology}, volume~2, pages
%  125--247. Academic Press, New York, 3rd edition, 1954.

%\bibitem[Keohane(1958)]{Keo:58}
%\newblock \emph{Power and Interdependence: World Politics in Transitions}.
%\newblock Little, Brown \& Co., Boston, 1958.

%\bibitem[Powers(1985)]{Pow:85}
%\newblock Is there a way out?
%\newblock \emph{Harpers}, pages 35--47, June 1985.

%\bibitem[Soukhanov(1992)]{Heritage:92}
%\newblock \emph{{The American Heritage. Dictionary of the American Language}}.
%\newblock Houghton Mifflin Company, 1992.

%\end{thebibliography}

%\appendix
%\section{Tela do Sistema IoT de Monitoramento de HM}    % Each appendix must have a short title.

\end{document}